\documentclass[fleqn]{aa}  

\usepackage{graphicx}
\usepackage{xcolor}
\usepackage{hyperref}
\usepackage{txfonts}
\usepackage{calligra}
\usepackage{calrsfs}
\usepackage[T1]{fontenc}
\usepackage[mathcal]{eucal}
\newcommand{\vectBeta}{\bf {\it \beta}}

\newcommand{\vectTheta}{\bf {\it \Theta}}
\newcommand{\vectAlpha}{\bf {\it \alpha}}
\newcommand{\vectR}{\bf {\it R}}

\begin{document}

   \title{A fast algorithm for simulating light curves of stars at extreme magnification affected by microlensing.}

   \titlerunning{Fast Light Curves for Microlensing}
  \authorrunning{Diego et al.}

   \author{J.M. Diego
         \thanks{jdiego@ifca.unican.es}
          }

   \institute{Instituto de F\'isica de Cantabria (CSIC-UC). Avda. Los Castros s/n. 39005 Santander, Spain
             }


 \abstract{
     We present a fast algorithm to produce light curves of distant stars undergoing microlensing near critical curves.  The need of these type of algorithms is motivated by recent observations of microlensing events of distant stars at high redshift and at extreme magnification factors. The algorithm relies on a low-resolution computation of the deflection field due to an ensemble of microlenses near critical curves, and takes advantage of the slowly varying nature of the deflection field to infer the magnification of the unresolved images. 
   }
   \keywords{gravitational lensing -- microlensing -- dark matter -- cosmology
               }

   \maketitle
%

\section{Introduction}
The observation of Icarus in 2018 \citep{Kelly2018} by the Hubble Space Telescope (HST) represented the beginning of a new branch of astrophysics: The study of individual stars at cosmological distances. Icarus, at $z=1.49$ behind the cluster MACS J1149, was the first individual star ever observed beyond redshift $z=1$ and pushed the record of the farthest star by a factor $\approx 200$. Icarus was observed thanks to the combined effect of a large macromodel magnification factor ($\mu \approx 500$) plus the boost given by a microlens, momentarily intersecting the path of the photons (adding an extra factor $\mu \approx 4$ to the total magnification), resulting in a net magnification factor of $\mu \approx 2000$. At magnification factors of $\mu=2000$, there is a net gain of more than 8 magnitudes, making stars that would be undetected with current technology bright enough to be detected by telescopes like HST. This technique allows us to reach a vast volume beyond redshift $z=1$ comprising thousands of Gpc$^3$. This volume is millions of times larger than the volume accessible in our local universe without the aid of lensing, and where we have been so far limited to study a relatively small number of very bright stars.  
Stars undergoing microlensing events can be identified by algorithms designed to search for transients such as supernovae (SNe). Other stars that are being strongly magnified near critical curves have been identified in recent years. Behind the galaxy cluster MACS J0416, several candidate strongly lensed stars have been identified \citep{Rodney2018,Chen2019,Kaurov2019}. Among these, one candidate in particular, dubbed Warhol at $z=0.94$, has unambiguously identified as a star \citep{Chen2019,Kaurov2019}. Like in the case of Icarus, Warhol shows temporal variations probably due to microlenses. More recently, behind the galaxy cluster PSZ1 G311, \cite{Diego2019} identified an extremely magnified star dubbed Godzilla and at redshift $z=2.37$. As opposed to Icarus and Warhol, Godzilla was not identified due to temporal variations in the flux but due to its unusually bright appearance, lack of counterimages, and unresolved nature. Although a previously known object \citep{Vanzella2020}, Godzilla was originally unrecognized as an extremely lensed star. The high brightness in Godzilla is interpreted in \cite{Diego2022} as due to the fact that the star is being observed during an outburst phase (these outbursts can last decades in the observer frame). Due to its brightness ($m_{AB}\approx 22$), Godzilla represents the first star at $z>1$ for which a spectrum was obtained \citep{Vanzella2020}. The MUSE spectrum shows peculiar features also present in massive luminous blue variable stars like a P-cygni profile \citep{Diego2022}. 
An interesting fact regarding Godzilla is that no microlensing flux fluctuations have been observed so far. This is consistent with the magnification of the star being in the range of several thousands. In this situation, as discussed in \cite{Diego2022} and \cite{Welch2022}, there is an effect known as the "more-is-less" effect, where increassing the magnification results in more overlapping microcaustics in the source plane. Crossing one of these microcaustics results in a relatively small change in flux, since the flux linked to other microcaustics is still significant, and remain more or less constant during the microcaustic crossing. 

The latest example of an extremely magnified star is also the most distant one. \cite{Welch2022} reports the discovery of Earendel, a strongly magnified star at a record breaking redshift $z=6.2$, peering into the end of the  reionization epoch. Although unlikely to be a Pop III star, Earendel represents one of the missing links in the evolution of stellar populations. Earendel was discovered behind the galaxy cluster WHL0137–08 ($z=0.566$) in the strongly lensed galaxy Sunrise. This galaxy was first observed in images from the RELICS program \citep{Coe2019}. Although spectral measurements of Earendel are still not available, and photometric measurements are relatively poor, it is believed that Earendel is a star with very low metallicity and very luminous. Earlier estimates in \cite{Welch2022} would put Earendel in pair with the most luminous stars known in our local volume (${\rm R} < 40$ Mpc). As in the case of Godzilla discussed above, no significant flux variations are appreciated during the two epochs (3 years apart) where Earendel was observed. This can be explained by the same "more-is-less" effect since the estimated magnification of Earendel is also in the range of thousands, resulting in multiple overlapping microcautics. However, small flux fluctuations ($\Delta m \approx$0.5--1 magnitudes) are expected due to microlensing but one needs to wait for deeper observations with JWST\footnote{A JWST program to observe Earendel is already approved}. \\

The recently launched JWST telescope arrives at the perfect moment to exploit the recent discovery of extremely magnified distant stars. With its gold-coated mirrors, JWST will lead us to a new golden age in astronomy, exposing previously unknown distant objects with unprecedented detail. Some of these objects will undoubtedly include extremely magnified stars, discovered through the same technique so successfully proven by Hubble.  
As mentioned earlier, observation of these distant stars represents the beginning of a new branch of astrophysics where we will learn not only about the physics of the first and second generation stars, but also about the small scale fluctuations in the lens plane that perturb the magnification of the background stars. An interesting aspect of extremely magnified distant stars is that the ubiquitous presence of microlenses prohibits magnification factors much larger than a few tens of thousands, even for relatively compact stars with sizes of several solar radii \cite{Venumadhav2017,Diego2018}. This is interesting since it imposes a natural bias toward brighter stars. Without this limitation, there would exist a degeneracy between the more common stars that are fainter but more magnified and stars that are more luminous but less magnified. A maximum value for the magnification then sets a limit to the lowest possible luminosity of the star at a given redshift. Hence, JWST will naturally pick up the brightest stars at high redshifts, offering a clean view of the massive end of the primeval stellar mass function.\\

Microlensing events of these distant stars can provide a unique type of information about the components of the gravitational lens effect. One type of microlens, stars responsible for the intracluster light (or ICL), will certainly have an impact on the magnification of distant stars. The role of these microlenses has been extensively studied in the context of quasar microlensing \citep{Chang1979,Irwin1989,EstebanGutierrez2022}, but given the much larger size of quasar accretion discs, the role of stellar microlenses is significantly much smaller than in the case of microlensing of distant stars. A background star can act as a pencil beam scanning through the small scale anisotropies in the lens plane. The smaller the pencil beam (that is, the smaller the star radius), the better the resolution one can obtain in the lens plane. Although challenging due to the ubiquitous presence of the more massive stellar microlenses, with good cadence it would be possible to detect even rogue planets at the redshift of the lens \citep{Diego2018}, which are not orbiting a star. High-cadence observations of very bright stars, such as Godzilla, could potentially reveal the first planet at cosmological distances ($z>0.4$). \\

Future studies of lensed distant stars will have to rely on accurate light curves to interpret the results. For instance, since the maximum magnification during a microcaustic crossing depends on the radius of the star, these events can be directly used to set constraints on the radius of the star, and hence its surface gravity. Since massive stars usually form binaries, the light curve may reveal the existence of the binary since each star would cross a microcaustic a different times. Frequent monitoring of the system can be used to also derive orbital parameters of the system, since the relative distance between the stars in relation to the microcaustics will depend on the orbital phase. \\

For these studies we need accurate and fast computation of light curves with microlensing events that can later be compared with actual observations. Although similar tools already exist in the literature, they have often been applied to regimes where the magnification from a macromodel is small or negligible (for instance in microlensing within our own Galaxy), or the magnification is moderate, usually in the range of a few to a few tens, such as in the case of quasar microlensing. 
When considering extreme magnification one needs to consider magnification factors of order $\mu\sim1000$. Also, since one wants to simulate light curves that extend several decades, the simulation needs to be large enough in order to contain the path of a moving star through a web of microcaustics during that period. One can get a sense of the requirements from standard ray tracing approaches using simple estimates. Adopting a typical relative velocity of 1000 km s$^-1$ between the moving star and the web of microcaustics, a distant star would move $\approx 6$ microarcseconds through the web of microcaustics in five decades. If one wants to resolve a star with $\approx 10$ R$_{\odot}$ or $R\approx 3\times 10^{-5}$ microarcsec, the number of pixels in the source plane needs to be of order ${\rm N} \sim 10^{10}$. Since we are considering magnifications factors of order $\mu \sim 1000$, the corresponding number of pixels in the image plane needs to be ${\rm N} \sim 10^{13}$. The disk space for the deflection field would require $\sim 1$ Pb. Although technically doable, the computing resources for simulating just one light curve would be very demanding.  Alternative more efficient approaches have been devised over the last years that can cope with the simultaneous demand of large area in the image plane and high resolution in the source plane. In \cite{Venumadhav2017} the authors used an approach that exploits ideas from earlier work by \cite{Paczynski1986,Witt1993} to track the microimages forming near critical curves. In a more recent work, \cite{Meena2022} adopts ideas similar to the ones used in zoom-in N-body simulations to increase the resolution around lensed images. Both approaches produce robust light curves at high resolution and allows for efficient computation of the light curves. \\ 

In this work we present an approach that differs from the ones listed above. This method was originally employed to produce the light curves that were used to interpret the Icarus microlensing event in \cite{Kelly2018} and \cite{Diego2018}. With some delay over this earlier results, (but still consistent with the time dilation from stars at $z \approx 2$), we present in the next sections the method used in \cite{Kelly2018} and \cite{Diego2018}. 

\section{An efficient algorithm for light curves at extreme magnifications}\label{Sect_Algorithm}
The algorithm presented in this work takes advantage of the smoothness of the deflection field and interpolates this field to rapidly find the multiple images and their magnifications. In the next subsections we discuss the basic equations from lensing theory and present cartoon examples that help understand the main ideas behind the algorithm. 

\subsection{Formation of microimages around microlenses}
The lens equation is defined in the canonical way
\begin{equation}
\vectBeta = \vectTheta - \vectAlpha(\vectTheta)
\end{equation}
where $\vectBeta$, and $\vectTheta$ are angular positions in the source and image plane respectively, and $\vectAlpha$ is the deflection field. Since these fields are all vector fields, we can decompose them into their cartesian components. In particular we are interested into the two residual fields, $\vectR_x$ and $\vectR_y$ defined as;
\begin{eqnarray}
\vectR_x = \vectBeta_x - (\vectTheta_x - \vectAlpha_x(\vectTheta))  \\
\vectR_y = \vectBeta_y - (\vectTheta_y - \vectAlpha_y(\vectTheta)) \nonumber
\label{eq_R}
\end{eqnarray}
where the $x$ and $y$ subscripts represents the $x$ and $y$ components of the fields. \\

If we consider a small source, for instance a star, with radius ${\rm R}_*$, then images form at those locations where the following condition is met;
\begin{equation}
d=\sqrt{\vectR_x^2 + \vectR_y^2} < {\rm R}_*
\label{eq_d}
\end{equation}
Images in the image plane can be found very efficiently through minimization algorithms by taking advantage of the quadratic nature of $d^2$, which quickly converges to a valley of positions where distances are below some threshold in $d$. This threshold is typically set to a value larger than a few times ${\rm R}_*$. A second step follows which precisely finds the number and sizes of all microimages at the subpixel level. \\ 

In general, at extreme magnification values ($\mu ~1000$) near a galaxy or cluster critical curve, the deflection field is a contribution of a macromodel which produces a deflection field (with slope close to 1 in the direction of maximum magnification) and very small perturbations to this field from the microlenses near the critical curve. The deflection field from microlenses scales as $r^{-1}$ with $r$ being the distance to the microlens. Hence the perturbations from microlenses are also smooth, with the exception of very small distances $r<<r_E$, where $r_E$ is the Einstein radius of the microlens. However, at this very short distances  the magnification factors are very small ($\mu < 1$) and can be safely ignored. 

\begin{figure} 
   \includegraphics[width=9cm]{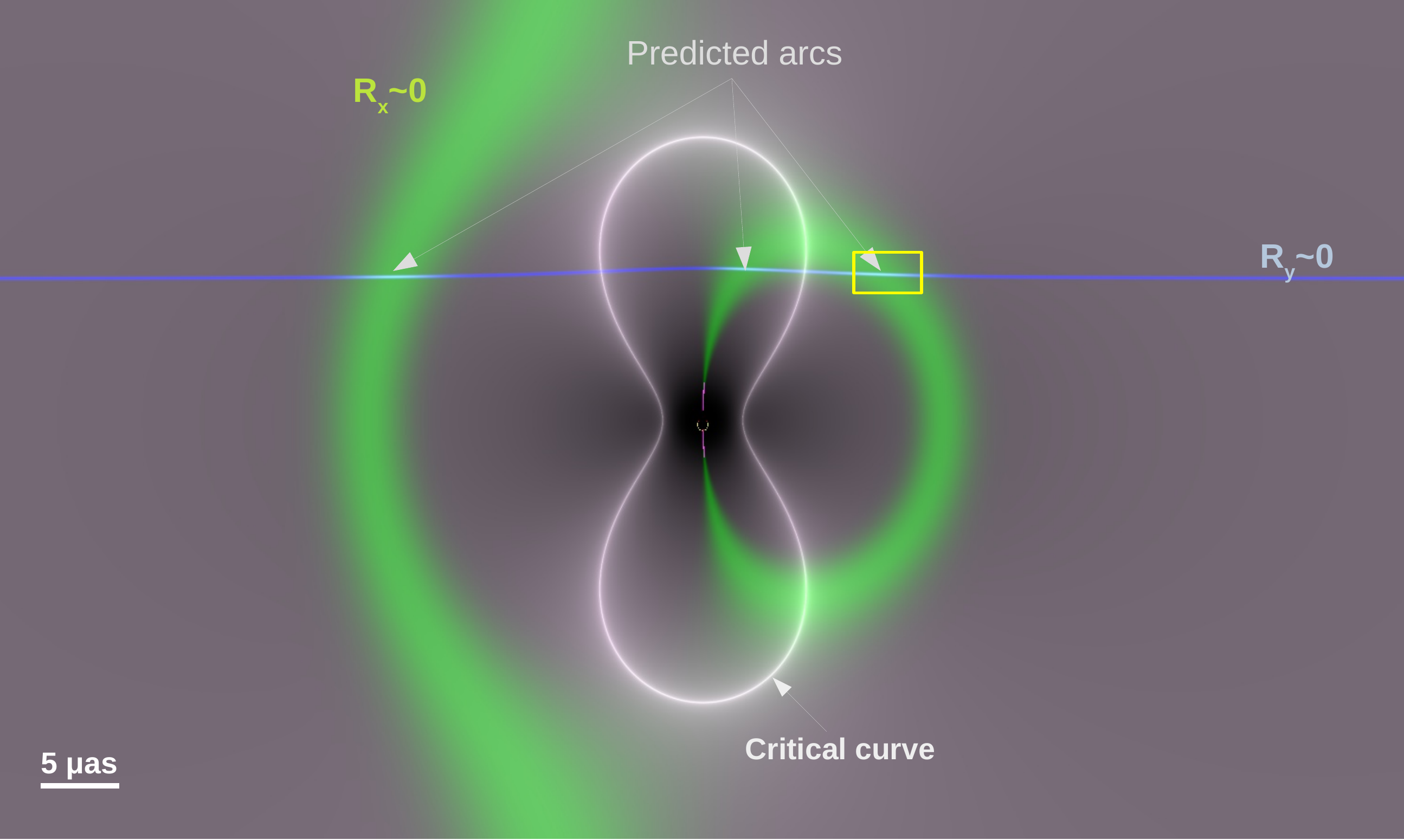}
      \caption{Magnification map around a 1 ${\rm M}_{\odot}$ embedded in a macrolens potential with magnification $\mu_t=100$ and $\mu_r=2$. The grey scale shows the magnification with the critical curve around the microlens forming a standing hour glass figure. The blue near horizontal thin curve marks the region where the residual $R_x$ is nearly zero. The green wider curve marks the corresponding region where the residual $R_y$ is close to zero. Arcs form in regions where both blue and green curves overlap, that is, when the condition in equation~\ref{eq_d} is satisfied. The yellow rectangle encloses a region similar to the cartoon representation of Figure~\ref{Fig_Cartoon1}. 
         }
         \label{Fig_OneMicrolens}
\end{figure}

Figure~\ref{Fig_OneMicrolens} shows a simple example of a microlensed image and how it relates to the residuals $R_x$ and $R_y$. For this particular case, the macromodel magnifies the source by a factor $\mu = \mu_t\times\mu_r = 100\times2 = 200$. Without the microlens there would be one single arc 200 times larger than the original source. For illustration purposes, in this figure we adopt a source that is large enough so the arcs can be appreciated visually. In a real situation where the source represents a star, the lensed microimages would in general be much smaller than the pixel size. In particular, the pixel size used for figure~\ref{Fig_OneMicrolens} is 30 nanoarcseconds which is orders of magnitude larger than the typical size of a luminous star at cosmological distances. 
The source is placed at $z=1.5$ and is modelled as Gaussian with dispersion 30 nanoarcsec. The images are simply given by $exp(-d^2/(2\sigma^2)$, where $d$ is given by~\ref{eq_d}. In the presence of the microlens (a star with 1 solar mass at redshift $z=0.55$), and for the particular position of the source in relation to the microlens,  3 microimages are formed (marked with arrows in the figure). The total magnification of the microimages is $\mu=410$. That is, the microlens boosts the magnification by a factor $\approx 2$. The critical curve around the microlens is the vertical hour-glass shape. The grey scale in the plot shows the magnification. The green and blue bands show regions where the residuals $R_x$ and $R_y$ are close to zero. In particular, these bands are given by $exp(-R_x^2/(2\sigma^2)$, and $exp(-R_y^2/(2\sigma^2)$ respectively. The product of these two bands is equal to $exp(-d^2/(2\sigma^2)$, that is, it results in the three microimages. The important result from this figure is to realize how both $R_x$ and $R_y$ vary slowly, and they can be approximated by straight lines once we consider scales comparable to the pixel scale. Under this condition, the behaviour of $R_x$ and $R_y$ can be easily predicted at scales smaller than the pixel scale, creating the foundation for the algorithm presented in the next subsection. 

\subsection{Interpolating at the subpixel level}
The previous section showed the simple example of a microlens producing three microimages. The pixels potentially containing microimages can be rapidly found by restricting the search to pixels where $d^2$ is below some threshold $\epsilon$, for instance $\epsilon=100R_*$ or more generally $\epsilon <1$ pixel, that is a fraction of a pixel. The sample of pixels that meet this requirement needs to be updated only a few times, since changes in $d$ are very small when the source position is varied by a distance comparable to $R_*$.
Restricting the search of microimages to the subset of pixels that satisfy $d<\epsilon$, and given the smooth (and linear) dependency of $d$, $R_x$ and $R_y$ with the source position, the small subsample of pixels containing microimages can be easily identified since $R_x$ and $R_y$ must contain a zero, i.e, the quantity $R_x-R_y$ changes sign between two opposing edges in the pixel. This change of sign can be easily computed by interpolating neighboring pixels. We refer to the subsample of pixels that contain microimages as $P_c$. 
After the preselection of $P_c$ is performed, the last element that remains is the determination of the magnification of the microimages within these pixels. \\

\begin{figure} 
   \includegraphics[width=9cm]{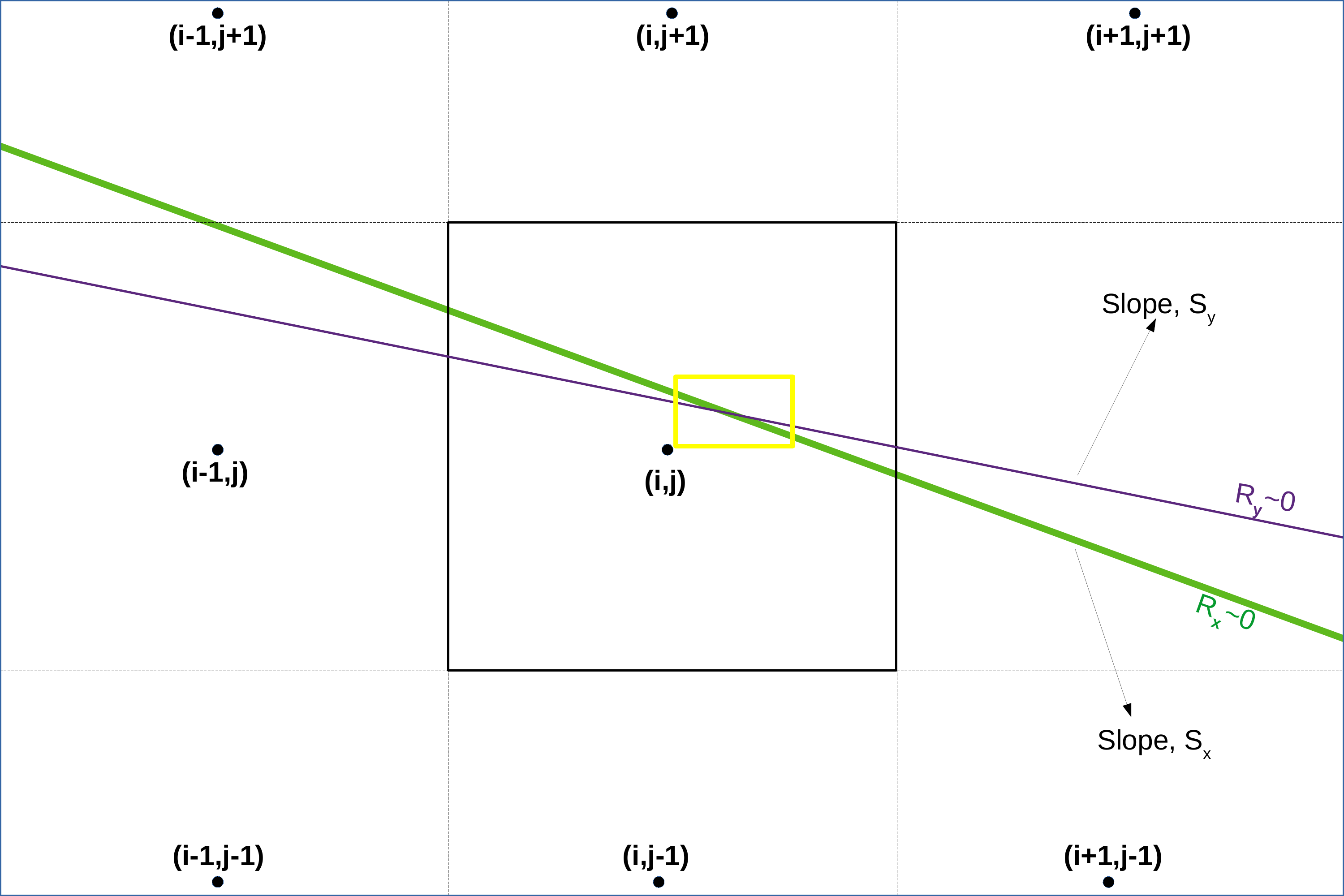}
      \caption{Representation of $R_x$ and $R_y$ at the subpixel area. The squares represent the pixel size. The pixel at position $(x,y)$, and containing the arc is marked with solid black lines in the center. Neighboring pixels are marked with dashed lines. The residual $R_x$ and $R_y$ are marked with blue and green lines respectively. The thickness of the line represents the points where $|R_x| = |R_y| = R_*$. The yellow rectangle is the region represented in greater detail in figure~\ref{Fig_Cartoon2}
         }
         \label{Fig_Cartoon1}
\end{figure}

Through a combination of linear interpolations it is possible to determine the size of the microimage within a pixel and hence the magnification. For circular sources such as stars, the intrinsic size of the source is given by $\pi R_*^2$ and the magnification of the microimage $i$ is given by $\mu_i=A_i/pi R_*^2$, where $A_i$ is the estimated area of the microimage $i$. Adding up the magnifications from all individual microimages we can compute the net magnification for that particular position, $(\beta_x, \beta_y)$, of the source.

\begin{figure} 
   \includegraphics[width=9cm]{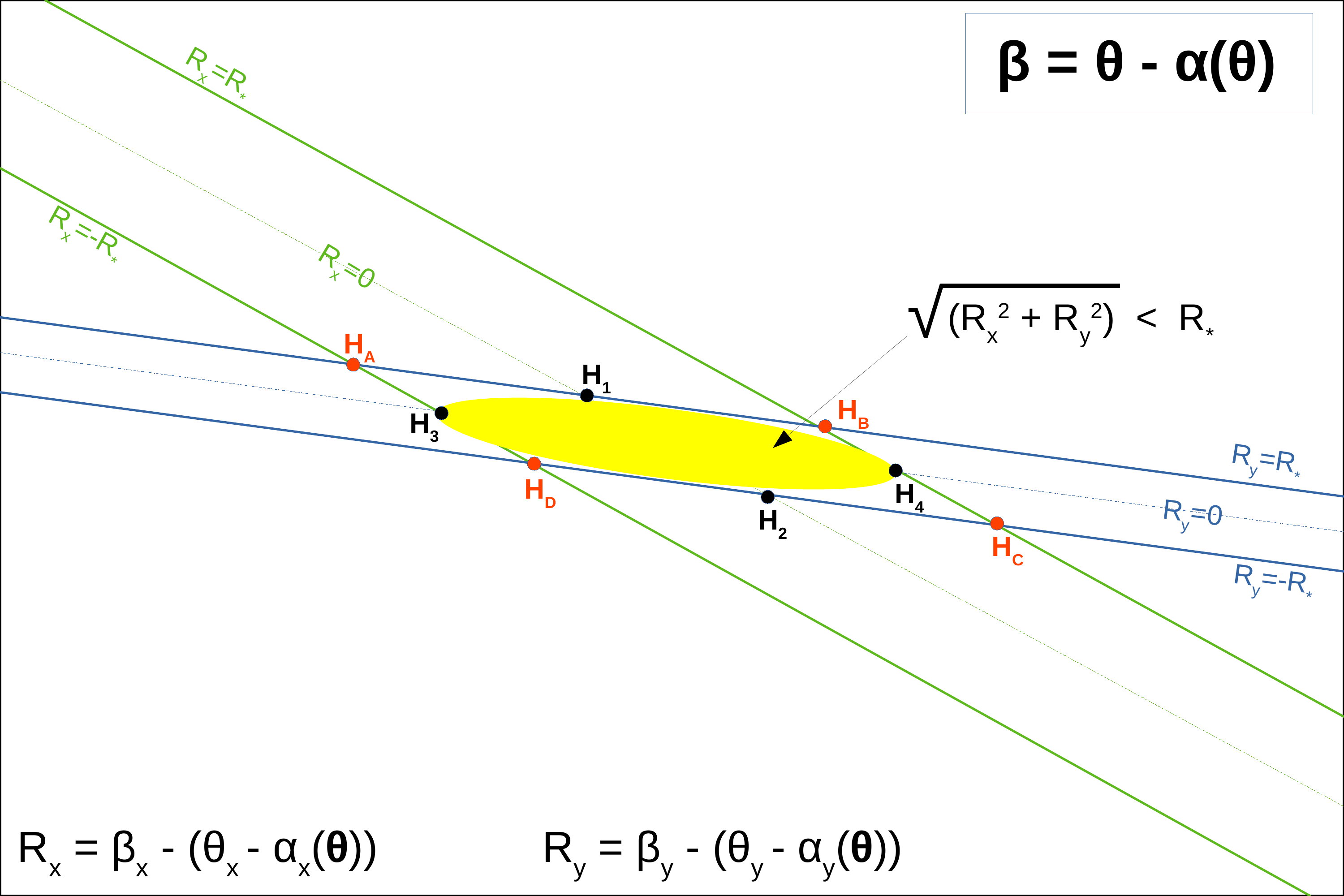}
      \caption{Cartoon representation of the algorithm. Points H$_1$ through H$_4$ are found through interpolation of the smooth ${\rm R}_{\rm x}$ and ${\rm R}_{\rm y}$ 2-dimensional fields, and determining the intersection points at ${\rm R}_{\rm x,y}=0,\pm{\rm R}_{*}$. The scale of the pixel is not shown but is assumed to be larger than the size of this figure (see figure~\ref{Fig_Cartoon1}).   
         }
         \label{Fig_Cartoon2}
\end{figure}

Taking advantage of the smoothness of the deflection field we can interpolate it to find the fraction of a pixel that contains a counterimage. A cartoon idea of the different steps in the algorithm is given in figures~\ref{Fig_Cartoon1} and \ref{Fig_Cartoon2}.

The size (or area $A_i$) of the microimage can be estimated by computing the 4 points ${\rm H}_A$, ${\rm H}_B$,${\rm H}_C$, and ${\rm H}_D$ shown in figure~\ref{Fig_Cartoon2} as the  intersecting points between the lines $R_x=\pm R_*$ and $R_y=\pm R_*$. 

Alternatively one can use the intersection points ${\rm H}_1$, ${\rm H}_2$,${\rm H}_3$, and ${\rm H}_4$. 

The magnification of each microimage is found by computing the ratio of areas. 
\begin{equation}
    A = \frac{A_p}{\pi R_*^2}
\end{equation}
Where $A_p$ is the area of the parallelogram defined by the intersection points ${\rm H}_A$, ${\rm H}_B$,${\rm H}_C$, and ${\rm H}_D$. This area needs to be corrected by the factor $\pi/4$ to account for the fact that the lensed image does not occupy the entire area within the parallelogram. The area $A_i$ is then given by; 
\begin{equation}
A_i = d({\rm H}_3,{\rm H}_4) \times {\rm h} \times \frac{\pi}{4}
\end{equation}

where $d({\rm H}_3,{\rm H}_4)$ is the distance between these points, and ${\rm h}$ is the distance between the two lines $R_x=R_*$ and $R_x=-R_*$. This distance is given by ;
\begin{equation}
h = \frac{|y_1-y_2|}{\sqrt{1+S_x^2}}    
\end{equation}
where $S_x$ is the slope of the line $R_x=0$ and $y_1$, $y_2$ are constants in the relation $y = y_{1,2} + S_x\,x$ which are determined through interpolation. The point $y_1$ is obtained as the point where $R_x=R_{*}$ between pixels (i,j+1) and (i,j-1). Similarly, the point $y_2$ can be obtained as the position where $R_x=-R_{*}$ between the same pixels.

\section{Comparison with ray-tracing}\label{Sect_Results}
\begin{figure} 
   \includegraphics[width=9cm]{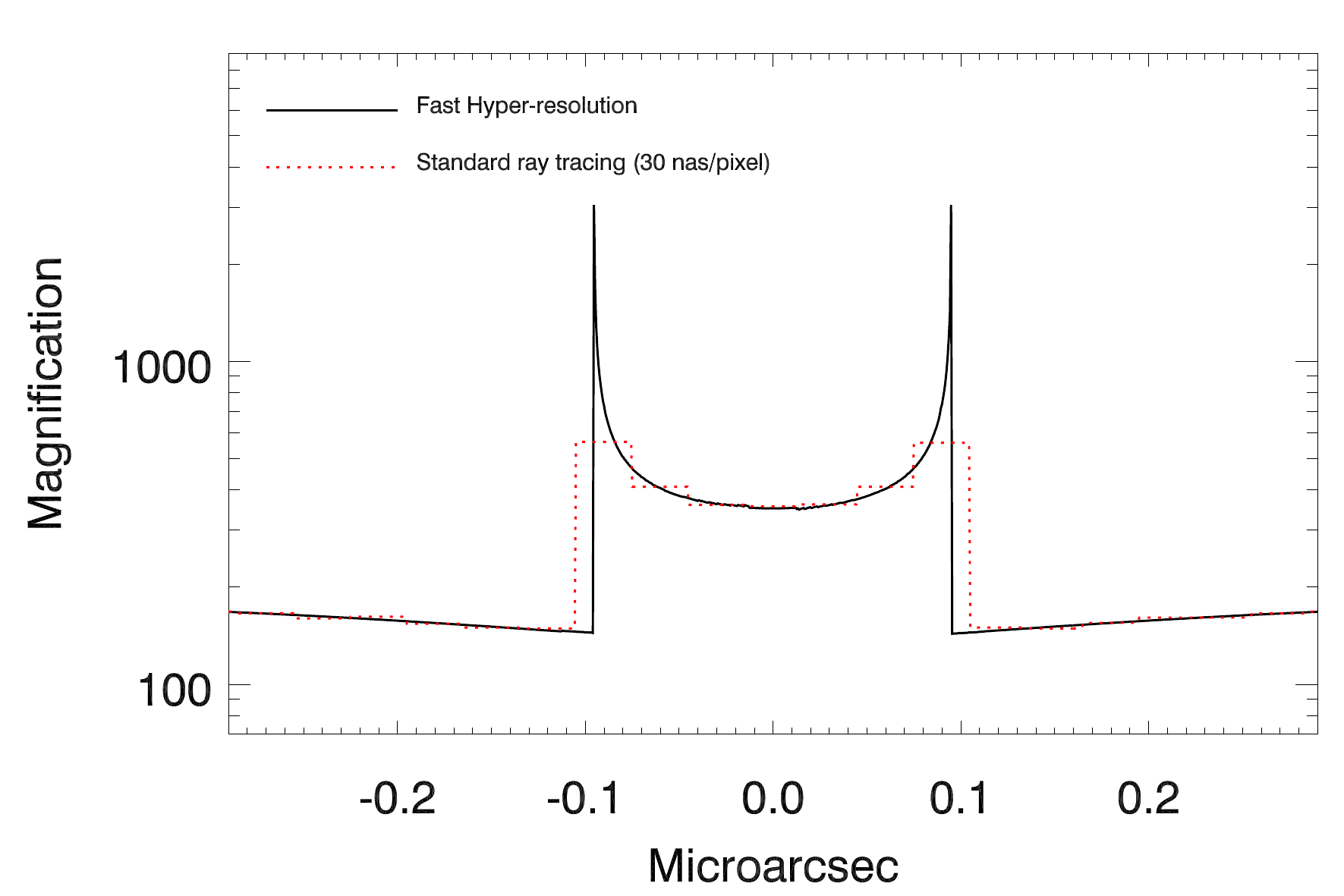}
      \caption{Comparison with standard ray shooting. The solid line shows the result from our method for a star with $R=20\, R_{\odot}$. For comparison we show as a red dotted line the result obtained from the ray-shooting method, with a pixel scale similar to the one used for the simulation in the image plane (30 nanoarcseconds). The trajectory of the star intersects the microcaustic horizontally. The microlens has one solar mass and microimages form along the curve $R_y=0$ shown in figure~\ref{Fig_Cartoon1}.
         }
         \label{Fig_FastLightCurve1}
\end{figure}

In order to test the performance of the algorithm we compute the deflection field with a pixel scale of 30 nanoarcsec per pixel and place a microlens in the centre with 1 $M_{\odot}$ (see figure~\ref{Fig_OneMicrolens}). For the background star we adopt a radius of 20 $R_{\odot}$, that is, the star is $\approx 2500$ times smaller in area than the pixel. The star moves in a horizontal direction with respect to the microlens shown in figure~\ref{Fig_OneMicrolens} and forms microimages along the $R_y=0$ region shown in the same figure. We compute the net magnification at each source position by adding the magnification of each microimage using the algorithm described above. The result is shown in figure~\ref{Fig_FastLightCurve1} as a black solid line. In a regular laptop, it takes less than a minute to compute this light curve (1000 varying star positions in the source plane). For comparison we show as a red dashed line the result obtained from the brute force approach of inverse ray shooting, where pixel positions in the image plane are projected back into the source plane. The standard ray shooting technique can not resolve the caustics well and predicts smaller magnification values. In contrast, our fast algorithm is able to properly resolve the caustics even for sources as small as $R=20\, R_{\odot}$. \\

\begin{figure} 
   \includegraphics[width=9cm]{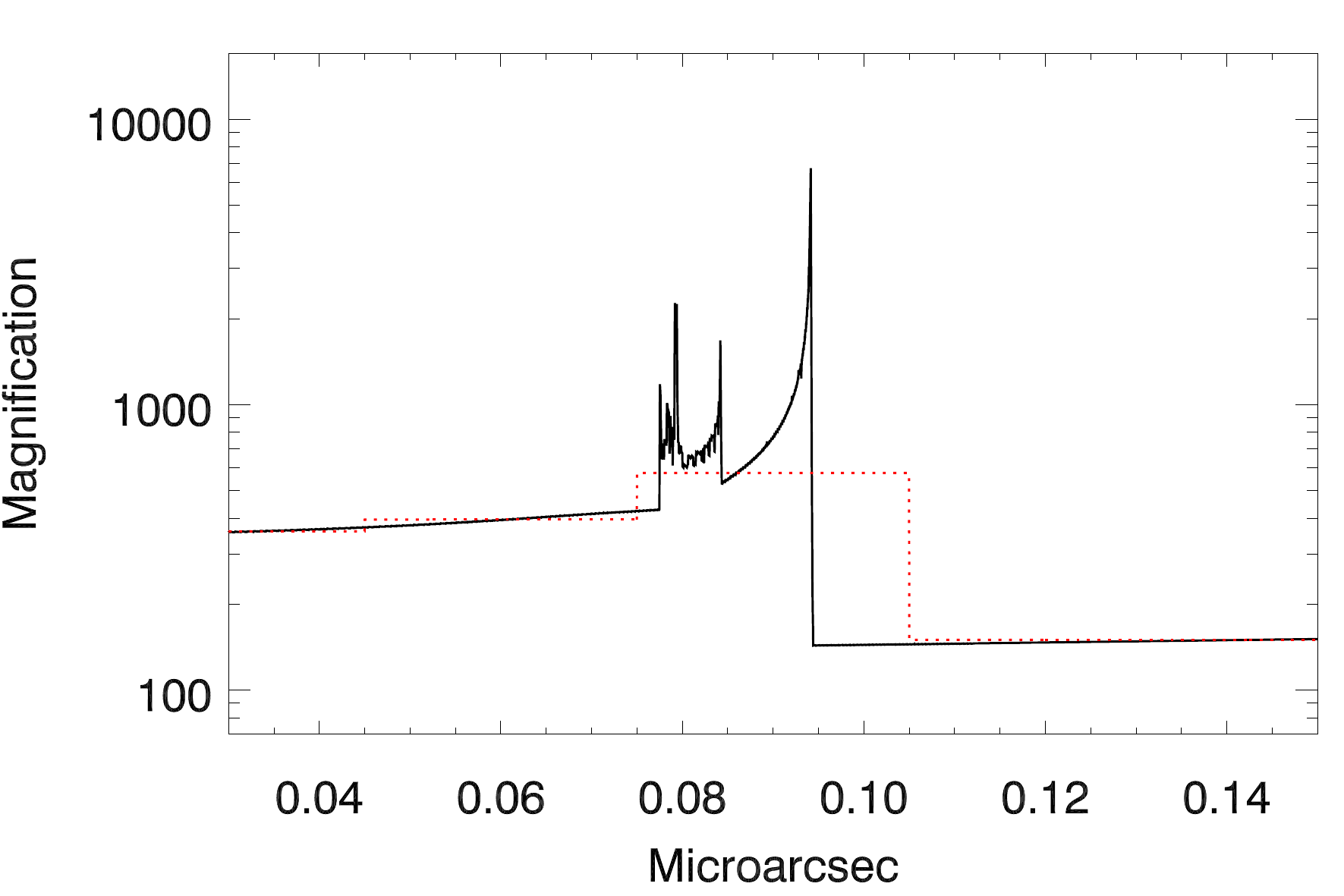}
      \caption{Zoom in near a caustic with a rogue planet  perturber. This plot shows a small portion of the critical curve for the same trajectory and microlens as in figure~\ref{Fig_Cartoon1} but where a small perturber with mass equal to one Jupiter mass intersects the trajectory of the counterimages near the right peak at $\approx 0.1$ microarcsec in figure~\ref{Fig_Cartoon1}. For this plot the background star is also smaller with radius $R=5\, R_{\odot}$. The red dotted lines shows the result obtained with the ray shooting method that does not resolve the small mass perturber. 
         }
         \label{Fig_FastLightCurve2}
\end{figure}

We also test the ability of the new algorithm to resolve small microlenses. We place a small microlens with one Jupiter mass (or $\approx 0.001 M_{\odot}$) near the caustic at $\approx 0.1 \mu$arcsec. In the image plane, this small microlens is at a physical distance of $\approx 0.1$ parsec from the larger microlens so it should be interpreted as a rogue planet not gravitationally bounded to the larger microlens. The critical curve of the Jupiter mass microlens is amplified by the combined magnification of the $1\, M_{\odot}$ microlens and the galaxy (or cluster) macrolens. 
In figure~\ref{Fig_FastLightCurve2} we show the light curve when the path of the background star intersects the small microcaustic from the rogue planet. For this particular case we have considered an even smaller background star with radius $R=5\, R_{\odot}$. The new algorithm is able to resolve both, the smaller Jupiter mass microlens and the smaller background star. As in figure~\ref{Fig_FastLightCurve2}, the red dashed line corresponds to the standard ray shooting technique that can not resolve the small microlens and fails at reproducing the magnification during the caustic crossing. The example shown in figure~\ref{Fig_FastLightCurve2} serves also to  illustrate the limitations of the new algorithm since small artifacts can be appreciated near the Jupiter mass microcaustic. In this case, the assumption made about the linear behaviour of the deflection field within the 30 nanoarcsec simulation pixel is not as accurate as in the case of the larger microlens. To better reproduce this case, a cubic interpolation or a smaller pixel size may be necessary. Nevertheless, this test is interesting to show that the algorithm can be used to efficiently produce light curves when a wide range of microlens masses are involved. Additional light curves can be found in the original paper that made use of this algorithm for the first time \citep{Diego2018}.

\section{Conclusions}\label{Sect_Conclusions}
We have presented a new algorithm to compute light curves of stars grazing a field of microcaustics near the caustic of a galaxy or cluster macrolens. In these scenarios, given the large magnification factors, $\mu_m$, from the macrolens (typically hundreds to thousands), the simulated region in the image plane needs to be relatively large. In the source plane, this large region gets compressed by the factor $mu_m$. If a luminous star is moving at a typical relative velocity of $\sim 1000 $ km s$^{-1}$ with respect to the micro-caustic network, it can cover $approx 1$ microarcsecond in the source plane during 10 years. The corresponding microimages would span a region $\mu_m$ times larger in the image plane. This means that in realistic scenarios the simulated region must cover approximately 1 milliarcsecond in the image plane. If one wants to resolve the magnification of stars (with typical sizes $\sim 10^{-11}"$) with standard ray shooting techniques the number of pixels to be simulated would be prohibitive. In this work we present an algorithm that relies on the smoothness of the deflection field in order to resolve microimages at the sub-pixel level. Using simulations with a pixel size that is orders of magnitude larger than the size of the stars, we show how it is possible to quickly produce light curves of a star moving across a web of microcaustics. The method is fast (less than one minute in a regular laptop for 1000 positions in the source plane) and accurate. We show how stars at cosmological distances as small as $5\, R_{\odot}$ can be resolved when the original simulation is done with a 30 nanoarcsecond pixel. We also show how small microlens with masses comparable to planets can be also resolved by the algorithm.

\begin{acknowledgements}
 J.M.D. acknowledges the support of project PGC2018-101814-B-100 (MCIU/AEI/MINECO/FEDER, UE) Ministerio de Ciencia, Investigaci\'on y Universidades.  This project was funded by the Agencia Estatal de Investigaci\'on, Unidad de Excelencia Mar\'ia de Maeztu, ref. MDM-2017-0765. 
 
\end{acknowledgements}

\bibliographystyle{aa} 
\bibliography{MyBiblio} 


\end{document}